\documentclass[default,iicol,pdflatex]{sn-jnl}% Default with double column layout

\usepackage{pifont}
\usepackage{caption}
\usepackage{subcaption}
\usepackage{tabularx}
\usepackage{multirow}
\usepackage{tablefootnote}
\usepackage{threeparttable}
\usepackage{listings}
\usepackage{multicol}
\jyear{2022}%

\theoremstyle{thmstyleone}%
\theoremstyle{thmstyletwo}%

\theoremstyle{thmstylethree}%

\begin{document}

\title[Article Title]{Combining Static Analysis and Dynamic Symbolic Execution in a Toolchain to detect Fault Injection Vulnerabilities}

%%=============================================================%%
%% Prefix	-> \pfx{Dr}
%% GivenName	-> \fnm{Joergen W.}
%% Particle	-> \spfx{van der} -> surname prefix
%% FamilyName	-> \sur{Ploeg}
%% Suffix	-> \sfx{IV}
%% NatureName	-> \tanm{Poet Laureate} -> Title after name
%% Degrees	-> \dgr{MSc, PhD}
%% \author*[1,2]{\pfx{Dr} \fnm{Joergen W.} \spfx{van der} \sur{Ploeg} \sfx{IV} \tanm{Poet Laureate} 
%%                 \dgr{MSc, PhD}}\email{iauthor@gmail.com}
%%=============================================================%%

\author*[1,2]{\fnm{Guilhem} \sur{Lacombe}}\email{guilhem.lacombe@cea.fr}

\author[2]{\fnm{David} \sur{Feliot}}\email{david.feliot@cea.fr}

\author[1,3]{\fnm{Etienne} \sur{Boespflug}}\email{etienne.boespflug@univ-grenoble-alpes.fr}

\author[1,3]{\fnm{Marie-Laure} \sur{Potet}}\email{marie-laure.potet@univ-grenoble-alpes.fr}

\affil[1]{\orgname{Université Grenoble Alpes}, \orgaddress{\city{Grenoble}, \country{France}}}

\affil[2]{\orgdiv{LETI CESTI}, \orgname{CEA}, \orgaddress{\city{Grenoble}, \country{France}}}

\affil[3]{\orgname{Verimag}, \orgaddress{\city{Grenoble}, \country{France}}}

%%==================================%%
%% sample for unstructured abstract %%
%%==================================%%

\abstract
{
Certification through auditing allows to ensure that critical embedded systems are secure.
This entails reviewing their critical components and checking for dangerous execution paths.
This latter task requires the use of specialized tools which allow to explore and replay executions but are also difficult to use effectively within the context of the audit, where time and knowledge of the code are limited.
Fault analysis is especially tricky as the attacker may actively influence execution, rendering some common methods unusable and increasing the number of possible execution paths exponentially.
In this work, we present a new method which mitigates these issues by reducing the number of fault injection points considered to only the most relevant ones relatively to some security properties.
We use fast and robust static analysis to detect injection points and assert their impactfulness. 
A more precise dynamic/symbolic method is then employed to validate attack paths.
This way the insight required to find attacks is reduced and dynamic methods can better scale to realistically sized programs.
Our method is implemented into a toolchain based on Frama-C and KLEE and validated on WooKey, a case-study proposed by the National Cybersecurity Agency of France.
}


%%\pacs[JEL Classification]{D8, H51}

%%\pacs[MSC Classification]{35A01, 65L10, 65L12, 65L20, 65L70}

\maketitle

\let\tmp\thempfn
\let\thempfn\relax
\footnotetext[0]{This work is partially supported by ANR-15-IDEX-02. All data generated or analysed during this study are included in this published article.}
\let\thempfn\tmp

\section{Introduction}\label{sec1}

From our credit cards to medical equipment and methods of transport, embedded systems are relied upon in nearly all aspects of modern life, including critical and sensitive applications.
Trust in these devices and their protective mechanisms is therefore paramount to ensure the viability of our professional endeavors and lifestyles.

One way the security of such devices can be ascertained is through certification processes such as Common Criteria \cite{common-criteria-1}.
This allows to gauge the difficulty of finding and performing attacks on the target system by measuring the time and level of expertise required to do this for a group of approved auditors leveraging state-of-the-art equipment and techniques \cite{JIL/19}.
Included in their arsenal are perturbation attacks, also known as fault injection attacks, which aim to cause exploitable faulty behavior in a system by subjecting it to extreme operating conditions.
Vectors of fault injection include applying strong electromagnetic fields to some components using a laser, causing power jolts or stressing DRAM memory via software to induce bitflips \cite{rowhammer,faults_survey}.
This can lead to secret information such as bits of cryptographic keys leaking through side channels \cite{faults_side_chan_aes} or even to loading an outdated firmware \cite{badfet17,VasselleTMME17}.

The tasks auditors must fulfill include reviewing the critical components of the target program such as cryptographic code, bootloaders or authentication procedures by exploring execution paths looking for dangerous behaviors, for example allowing to bypass said cryptographic code.
This latter task requires the use of tools which can have difficulties scaling to large programs due to path space explosion and infinite loops.
Moreover these issues only get worse when considering an active attacker who may influence execution by injecting faults.
Methods which could be used to reduce the analysis perimeter such as slicing are inadequate in this context as well.
Considering that auditors have limited time and knowledge of the code, fault analysis tools tend to be impractical on realistically sized programs.

To mitigate these issues, we propose an approach using static analysis to detect the most relevant fault injection points in a program relatively to security properties which an opponent may want to attack.
It involves finding faultable instructions in the dependencies of the properties and formally checking that they may have an impact.
This way the number of execution paths to consider is reduced to only those which cannot be formally proven harmless.
As a result more precise tools such as fault simulators, which generate mutated programs with simulated faults, and fault analysis tools based on dynamic symbolic execution \cite{Godefroid12,Cadar13}, which only generate a single mutated program for analysis, can be used on realistically sized programs more effectively.

In Section 2 of this paper we place our contributions within the current state of the art. 
We then present our method in Section 3, discuss some additional heuristics in Section 4 and provide experimental validation in Section 5.
Finally, we discuss related works in Section 6.

\section{Context and Contributions}

\subsection{Security Evaluation and Fault Analysis}

As part of the certification process, the auditors working for the Information Technology Security Evaluation Facilities (ITSEFs) must identify potential vulnerabilities to fault attacks in their target.
This \textit{fault analysis} either helps to discover actual exploits or to assert that the target is secure against some attacker models.
It also allows to find theoretical exploits which could not be performed within the limited time frame of the audit and would have been missed otherwise.

%When sources are available, auditors start from the source code and a state-of-the-art in terms of attack scenarios.
%From there they track logical vulnerabilities that can generally be replayed at the binary level \cite{RivierePLBCP14}.
When sources are available, auditors analyze them first in order to familiarize themselves with the target.
Potential attacks can be detected based on fault models (i.e. known high-level effects of hardware faults), giving insight on what parts of the code are likely to be vulnerable to fault injection as it has been shown that a combination of a powerful fault model and multi-fault analysis effectively covers low-level attacks \cite{RivierePLBCP14}.
This information can then be used by developers to decide where additional countermeasures are needed as well as to guide further evaluation at instruction and hardware levels.
Thus source-level fault analysis is an important first step in evaluating the resistance to fault injection attacks of a program despite being often overlooked.

Fault analysis is often associated with the evaluation of cryptographic implementations.
While fault attacks are a major threat in that context, another equally dangerous application of fault injection is to disrupt program logic outside of cryptographic components, sometimes bypassing them entirely.
For example, faults have been used to alter the control-flow of bootloaders and perform exploits on real hardware \cite{Bozzato}.

The main distinction between these two applications of fault injection is that the types of faults considered differ.
If the fault models chosen in a crytographic context tend to focus on altering data, for example by allowing to set values to zero, those used for evaluating non-cryptographic code focus more on altering control-flow as it is often more logic- and decision-centric rather than computation-heavy.
As a result the skillsets required to evaluate these two types of code are different and thus they are handled by separate teams.

In this work we focus primarily on the evaluation of non-cryptographic programs with complex control-flow resulting in a large set of possible execution paths, which grows exponentially when faults are taken into account.
However we also discuss its application in a cryptographic setting at the end of section 5.

\subsection{Motivating Example}

The following is a discussion of an example program which contains vulnerabilities to fault injection despite the presence of countermeasures.
We will also use this example to illustrate our fault analysis method later.

\begin{figure*}[t]
	\begin{lstlisting}
typedef struct data{
	§size_t§ msg_size;                  //input > 255
	char msg[256];                          //input
	§uint32_t§ key[8];                        //secret
} §data_t§;

void print_message(§data_t§ *d){
	unsigned int size = d->msg_size & 0xff; //countermeasure, fault 0x000000ff to 0xffffffff
	for(unsigned int i = 0; i <= size; i++}{
		if(i > size)                          //bypassed countermeasure
			return;
		//@ assert \valid_read(&d->msg[i]);   <- expected security property
		printf("%c", d->msg[i]);              //bytes of the key in the output!
	}
	printf("\n");
}
	\end{lstlisting}
	\caption{Example of a function vulnerable to fault injection}
	\label{fig:motex}
\end{figure*}

The function presented on Figure \ref{fig:motex} prints a message based on its size, both of which are controlled by the user.
Under nominal circumstances this is not an issue since a mask is applied to the size on line 8, limiting its maximum effective value to 255 and thus preventing buffer overflows.
The index is also checked to be within the expected bound of the buffer on line 10 in an attempt to thwart fault injection attacks.

However, attacking the countermeasure on line 8 by forcing the mask to 0xffffffff with a fault, which is a commonly considered outcome, results in the exact user provided size being used.
This also bypasses the index check on line 10.
In this example a secret cryptographic key is conveniently stored near the message in memory.
Inputting a greater than 255 value as the message size will therefore result in bytes of the key leaking in a similar way as with the Heartbleed OpenSSL vulnerability \cite{heartbleed} and violating the property expressed on line 12\footnote{In ACSL, \textit{valid\_read} allows to check that the content of pointers can be safely read as per the C standard.}.

Ignoring the fact that storing a secret key in such a fashion is inadvisable, detecting such vulnerabilities to fault injection can be difficult when they are buried deep within an application.
In fact, our example was inspired by an attack that was found on the ANSSI's WooKey project\footnote{See the 2020 Inter-CESTI challenge report \cite{inter_cesti}, section 9.} in a library comprised of roughly 2.5k lines of code \cite{code_iso}\footnote{See the SC\_get\_ATR function.}, which violated a similar property to the one on line 12, leading to a stack buffer overflow.
Additionally, the presence of some commonly used countermeasures may hide the issue to visual inspection.
The use of automated analysis tools would therefore allow to not only more reliably detect fault injection vulnerabilities, but also to evaluate the effectiveness of countermeasures.

\subsection{Usage of Tools in Security Evaluation}

As part of their job, security evaluators must discover vulnerabilities within a limited time frame while taking into account the current state-of-the-art.
Automated tools are therefore of interest to them as they allow to speedup analysis while offering specific guarantees in terms of coverage.
Another point of interest is that they allow to easily replay analyses, rendering cross-checking and updating results much more efficient.

Using such tools presents its own set of challenges however, as parameterizing them, creating attack scenarios and interpreting results all require extensive knowledge and experience to be done correctly.
One particular difficulty of interest to us is the definition of an analysis perimeter, which is an area of the target program containing relevant elements toward the chosen attack scenario that the analysis will be restricted to, with the intention of improving scalability.
This is especially important when analyzing large programs which tools may struggle to handle if considered in their entirety.
For example, evaluators may want to only analyze a few functions while stubbing the rest, with restrictions on inputs.

Ideally, a sound approach to extracting an analysis perimeter should be favored in order to ensure that no attack will be missed.
However this is often impractical when tools are suffering from scalability issues, which may cause the analysis to not terminate within a reasonable time frame or precision to be insufficient.
A solution is to progressively restrict the analysis perimeter based on heuristics, but since it can be difficult to judge the progress of analyses, determining when to interrupt them and apply further restrictions is a blind process.
For this reason, a perimeter widening approach consisting in progressively analyzing more code with less restrictions is often preferred as previous terminating analyses give a frame of reference for forming expectations on runtime.

In this context, using scalable but less precise static analysis techniques in order to guide the usage of more precise but less scalable dynamic ones is becoming common practice \cite{electronics}.

\subsection{Program Analysis Techniques}

Many program analysis methods and tools can be used for security evaluation.
The following is a succinct presentation of commonly used ones which are relevant to this work.

\paragraph{Abstract Interpretation}

Abstract Interpretation is a static analysis theory for the sound approximation of program semantics applied to ordered domains such as lattices.
It is used to gain some insight on the properties of the program without fully executing it.
Applications include computing value ranges for variables, data-flows and dependencies as well as checking assertions.
Implementations tend to scale well to realistically sized programs, although with stubbed functions and some loss of precision.

\paragraph{Dynamic Symbolic Execution (DSE)}

DSE is a dynamic analysis technique consisting in exploring execution paths in a program while constructing constraint formulas for variables with unspecified values, which are referred to as symbolic variables.
These constraints can then be solved using SMT solvers to check the feasibility of execution paths and to obtain inputs triggering them.
DSE implementations tend to have scalability issues when constraint formulas become too complex or due to path space explosion.
They can also get stuck in infinite loops and never terminate.

Concolic execution is a variant which allows to concretize symbolic variables (i.e. give them a concrete value) to mitigate these issues.
However this is not always enough and may induce a loss of correctness and completeness \cite{godefroid11}, i.e. some explored paths may not actually be feasible and some feasible paths may not be explored.

\subsection{Difficulties of Fault Analysis}
\label{sec:sli}

Fault analysis is particularly tricky as most widely used analysis techniques and tools are not designed with faults in mind.
The main issue is that faulty behavior must be accounted for at various program points, which we refer to as \textit{fault injection points}.
This introduces a large number of new feasible execution paths, which increases exponentially with the number of faults considered.
Thus the choice of the analysis perimeter becomes crucial as it directly impacts this issue.

Extracting an analysis perimeter is usually done by expressing assertions related to security properties and slicing \cite{slicing_survey,slicing_survey_2} based on dependencies.
This allows to produce a minimal program corresponding only to relevant execution paths with regard to the assertions.
The issue with this method is that faults may redirect control flow and induce paths which are normally unfeasible.
It could therefore result in the loss of attack paths, especially in a multi-fault context.

\begin{figure*}[t]
	\begin{lstlisting}
void process(int a, int b){ //function to analyze
	if(a && b){               //nominal path
		if(!a || !b) exit(1);   //countermeasure
		assert(a && b);
		...
		return;
	}
	if(a){                    //reachable by faulting the previous test
		if(!a) exit(1);         //countermeasure
		assert(a && !b);
		...
		return;
	}
	...
}

int analysis_main(){
	process(1, 1);
	return 0;
}
	\end{lstlisting}
	\caption{Program with an unfeasible execution path becoming feasible with a fault}
	\label{fig:unfex}
\end{figure*}

The example from Figure \ref{fig:unfex} shows a program setup for the analysis of the \textit{process} function.
In this case, both $a$ and $b$ are set to 1, which should result in the test on line 2 being always positive under nominal condition.
However a fault can be used to invert the result of this test, which would result in the execution reaching the one on line 8.
Since the countermeasure there does not check that $b$ is null, the assertion line 10 would be violated.
However this execution path would be lost after slicing since it would be detected as unfeasible by a precise dependency analysis, resulting in a potential attack being missed and illustrating the fact that this approach is not adapted for fault analysis.
Slicing may also result in countermeasures being removed since they often appear as redundant or dead code under nominal conditions.

Another issue with slicing is that code may be modified to the point that making the connection to the original is difficult.
This can make the interpretation of results difficult as well as resuce trust in the accuracy of the analyses.

The consequence of these difficulties is that reducing the size of the analyzed program in any meaningful way is often not possible in the context of fault analysis.
There is however another way in which we may effectively reduce the analysis perimeter, which is to reduce the number of fault injection points considered.

\subsection{Contributions}

\begin{itemize}
	\item 
We propose an approach using formal methods to find the injection points that a security property depends on and eliminate those which can be proven to have no impact. 
Faulty behavior is simulated using a generic fault model adapted to source-level analysis.
Relevant injection points are then selected for further analysis with a dedicated fault analysis tool.
	\item 
We implement our method as a toolchain based on proven and widely used tools which is suitable for both single- and multi-fault analysis.
The static analysis part is implemented as a Frama-C plugin.
Our fault analysis tool of choice is Lazart, which is itself based on dynamic symbolic execution by KLEE.
    \item
We present and test various heuristics allowing to deal with the scalability issues that arise when using dynamic symbolic execution for fault analysis.
	\item
We validate our method by using our implementation to find fault injection vulnerabilities in the ANSSI's WooKey secure USB storage device \cite{wookey}.
This results in weaknesses being discovered in the countermeasures of the bootloader part.
We also complement our experiments with an analysis of the verifyPIN program from the FISSC fault injection test suite \cite{fissc} and the \textit{sudo} unix command \cite{sudo}.
\end{itemize}

This paper presents multiple extensions of our original work.

\begin{itemize}
    \item 
We revised our method to streamline it, with a greater emphasis on our generic fault model.
Our implementation was also updated to use the latest version of Lazart.
    \item 
New selection heuristics designed to help with the scalability of analyses were added and tested.
    \item 
We tested our method on new targets from FISSC and validated our proposed fixes to WooKey's bootloader against single-fault attacks.
We also showed that these new countermeasures do not hold against double-fault attacks.
\end{itemize}

Our method allows to reduce the number of fault injection points considered in the fault analysis of a program by removing those which can be formally proven to have no impact on security properties.
The use of static analysis to this end allows to handle difficult execution paths and should result in better scalability when eliminating false positives using dynamic symbolic execution, as illustrated in section \ref{sec:res}.

\section{Our Method}

Our goal with this work is to design a method allowing to improve the scalability of fault analysis tools by reducing the number of injection points considered without risking to remove important parts of the target program nor losing attack paths.
We chose to use static source code analysis as a front-end in order to help with the evaluation of code-level countermeasures.
Additionally this helps to approach unknown code as the relevant parts to the targeted security properties can be easily pointed out.
Dynamic symbolic execution analysis then provides more precise information which is easier to interpret in this context.

\subsection{Tools}

We based our method and its implementation on proven and widely used tools, which need to be discussed as they impacted design decisions.
It is however important to note that the general concepts of our method should be applicable to any other tools.

\paragraph{Frama-C}
Frama-C \cite{framac} is a static analysis platform for the C language.
It parses .c files into a formal AST structure (based on CIL) which supports annotations detailing specifications and properties using the ACSL specification language \cite{acsl}.
Frama-C then manages plugins which implement various kinds of analyses.
We will be interested in the following ones:
\begin{itemize}
	\item \textbf{Eva \cite{eva,evaman}} performs abstract interpretation and computes abstract domains for variables in a program, including aliases. 
It can then use this information to correctly prove or disprove assertions expressed in ACSL, with inconclusive attempts being labeled as such.
Eva is thus useful not just for bug-finding but also for proving properties.
	\item \textbf{Pdg \cite{pdgman}} computes intra-procedural memory, data and control dependencies as dependency graphs. It is based on data-flow analysis using Eva, which allows for greater precision as values can be taken into account.
    \item \textbf{From} computes inter-procedural dependencies using data-flow analysis through Eva.
\end{itemize}

\paragraph{Lazart}
Lazart \cite{lazart,FDTC20} is a multi-fault analysis tool based on the KLEE concolic engine \cite{CadarDE08} which detects multi-fault injection attack paths.
This is achieved by mutating the program using a Clang compilation pass in order to simulate faulty behavior based on some fault models.
A dynamic symbolic execution analysis is then performed at LLVM IR level to find execution paths violating a security property. 
Correctness and completeness are inherited from KLEE, thus false positives may appear or attack paths may be lost if variables are concretized.
Lazart is a state-of-the-art tool with concrete use-cases in the context of fault analysis and security evaluation by ITSEFs \cite{inter_cesti}.

As previously discussed, dynamic symbolic execution suffers from scalability issues mainly due to path space explosion.
For this reason large programs with many potential fault injection points are difficult to analyze with Lazart, especially in a multi-fault context.
However since the static analyses implemented by Frama-c's plugins are based on abstract interpretation, they are less prone to these issues and can thus be used to mitigate those from Lazart.

\subsection{A Generic Fault Model}

Inferring the effect of hardware level faults on programs is very difficult as it would require a full understanding of the target platform from micro-architectural details to its firmware's implementation.
For this reason, most fault analysis techniques rely on fault models to determine faulty behaviors that may occur during execution.
Commonly considered fault models include the following:
\begin{itemize}
    \item \textbf{Data Load:} The value obtained when reading a variable can be altered (e.g. set to zero).
    \item \textbf{Test Inversion:} The outcome of a conditional jump can be inverted.
    \item \textbf{Control-flow Violations:} Instructions may be skipped and jumps may lead to unexpected locations.
\end{itemize}
These do not correspond to distinct types of faults but are rather higher level interpretations of faulty behavior which may overlap depending on scenarios.
For example, faulting the value of a test condition (data load model) is equivalent to inverting the test itself (test inversion model).
The data load and test inversion models are particularly relevant to us since they are available in Lazart.

Given our goal of evaluating at source level which potential fault injection points in a program are worth considering, we should choose a fault model generic enough to be relevant toward those previously described.
We should also take into account the limitations of source analysis, namely the lack of information on the runtime memory and the compiled binary layouts.
Our fault model should therefore be restricted to faulty behaviors which can be statically processed without this information.
For example, it should not allow to skip control instructions such as conditional jumps as what code would be executed next cannot be inferred at source level.
This is acceptable as such behavior would be extremely difficult to explore with current fault analysis tools.
Additionally some scenarios can be simulated in more source-friendly ways e.g. skipping function calls can be done by encapsulating them in tests.

%\begin{figure*}[t]
%    \begin{subfigure}{.45\textwidth}
%        \begin{lstlisting}
%...
%int i=0;
%while(i<10)
%{
%    ...
%    i=i+1;
%}
%...
%        \end{lstlisting}
%        \caption{While loop without faults}
%    \end{subfigure}
%    \hfill
%    \begin{subfigure}{.45\textwidth}
%        \begin{lstlisting}
%...
%int i=fault(0);
%while(fault(i<10))
%{
%    ...
%    i=fault(i+1);
%}
%...
%        \end{lstlisting}
%        \caption{While loop with expression faults}
%    \end{subfigure}
%	\caption{The expression fault model applied to a loop}
%	\label{fig:efex}
%\end{figure*}

\begin{figure*}[t]
	\begin{lstlisting}
typedef struct data{
	§size_t§ msg_size;
	char msg[256];
	§uint32_t§ key[8];
} §data_t§;

void print_message(§data_t§ *d){
	unsigned int size = àfault(d->msg_size àè&èà 0xff)à;
	for(unsigned int i = àfault(0)à; àfault(i <= size)à; i = àfault(i + 1)à}{
		if(àfault(i > size)à)
			return;
		printf("%c", àfault(d->msg[i])à);
	}
	printf("\n");
}
	\end{lstlisting}
	\caption{Expression fault model applied to our motivating example}
	\label{fig:efex}
\end{figure*}

We thus propose the following fault model:
\begin{itemize}
    \item \textbf{Expression Fault:} The evaluation of an expression can be altered. "Expression" refers here to \textit{exp} objects in the CIL AST, which correspond to C expressions as defined in the C standards minus assignments\footnote{In this case, only the right-side of the assignment is translated to \textit{exp}.}.
\end{itemize}
Additionally, we restrict our fault model to non-pointer expressions\footnote{Note that array indexes can still be faulted.}.
This, along with the fact that control instructions cannot be skipped, ensures that the control-flow graph of the program is never violated.

Figure \ref{fig:efex} shows how this fault model affects our motivating example.
Our fault model encompasses the effects of data load faults (lines 8, 9 and 12) and allows for test inversion since test conditions can be directly faulted (lines 9 and 10).
The effect of skipping instructions and faulting pointers on non-pointer data is also covered.
For example, skipping the loop counter incrementation on line 9 is equivalent to faulting the value of $i + 1$.
However these behaviors may require multiple faults to be covered in general e.g. to cover the effect of skipping a call to a function with implicit outputs (i.e. assignments to global variables, input pointer memory...) expressions containing them should all be faulted.

This model also encompasses faulty behavior which the others do not cover, such as errors occurring during computation of operations, e.g. obtaining an odd result from a multiplication by two.

\subsection{Overview of our Method}

\begin{figure*}[t]
	\centering
    \includegraphics[width=\textwidth]{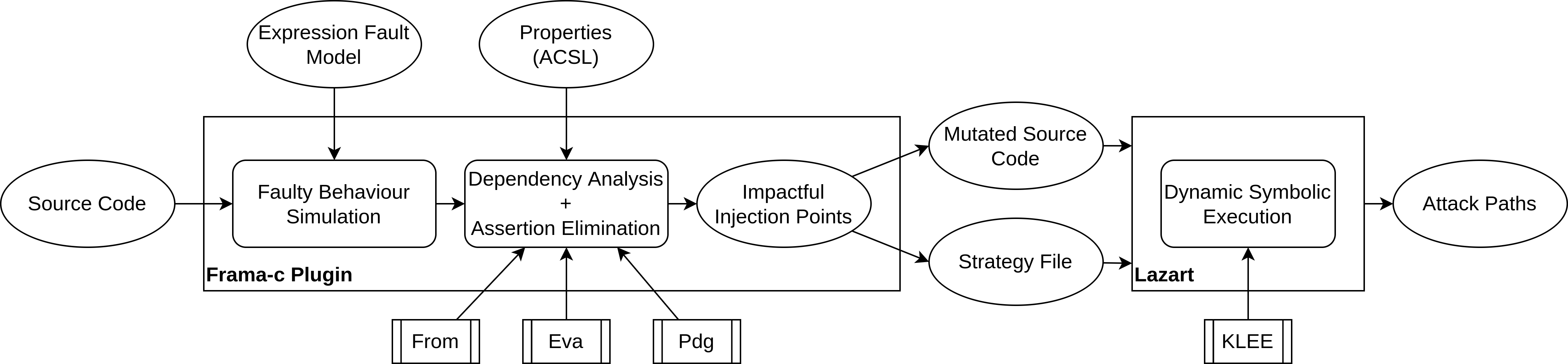}
	\caption{Overview of our method}
	\label{fig:workflow}
\end{figure*}

Figure \ref{fig:workflow} gives an overview of our method, which takes a source file with security properties expressed in ACSL as input.
The static analysis part is performed automatically by our Frama-C plugin and its outputs are immediately ready to be analyzed with lazart.

While properties can be difficult to extract from unknown code, the knowledge required to express them cannot be greater than that required to find attacks.
This is assuming that an attacker would look for the most relevant security properties to violate rather than proceed blindly when attempting to attack a system, making the definition of properties a requirement for finding attacks.

Security properties can be defined manually by adding ACSL assertions to the potentially vulnerable code, however this task can be challenging if many must be introduced or if it is not obvious which ones are relevant.
These difficulties can be mitigated by using automated tools to generate assertions or requiring developers to provide a formal specification of their programs.
For example, Frama-c's RTE plugin can be used to automatically add assertions for the purpose of discovering runtime errors such as buffer overflows.
Another plugin, METACSL, allows to generate assertions based on high-level, user-defined properties \cite{metacsl}.

\subsection{Introducing Faulty Behavior}

As faulty behavior must be taken into account when analyzing programs, the first step we have to take is to simulate it everywhere indiscriminately.
Then we will be able to conduct further analysis and select the most important injection points.
As Eva is the basis for all of the Frama-c analyses we will rely on, we thus need to find a way to force value ranges computed for expressions to be imprecise in order to enforce our fault model.

%\begin{figure}[h!]
%	\begin{lstlisting}
%extern int a,b,c;
%...
%int i=0^a;
%while((i<10)^b)
%{
%    ...
%    i=(i+1)^c;
%}
%...
%	\end{lstlisting}
%	\caption{Example of simulated faults on a loop}
%	\label{fig:xor}
%\end{figure}

\begin{figure*}[t]
    \begin{lstlisting}
extern int fault_5;
extern unsigned int fault_4;
extern int fault_3;
extern int fault_2;
extern size_t fault_1;
extern unsigned int fault_0;

void print_message(data_t *d_0)
{
  /*@ assert rte: mem_access: \valid_read(&d_0->msg_size); */
  size_t size = (d_0->msg_size & (unsigned int)0xff) ^ àfault_4à;
  {
    size_t i = (unsigned int)0 ^ àfault_0à;
    while ((i <= size) ^ àfault_2à) {
      if ((i > size) ^ àfault_3à) goto return_label;
      /*@ assert rte: index_bound: i < 256; */
      /*@ assert rte: mem_access: \valid_read(&d_0->msg[i]); */
      printf("%c",(int)d_0->msg[i] ^ àfault_5à);
      i = (i + (size_t)1) ^ àfault_1à;
    }
  }
  printf("\n");
  return_label: return;
}
    \end{lstlisting}
    \caption{Motivating example with simulated faults}
    \label{fig:xor}
\end{figure*}

As shown on Figure \ref{fig:xor}, this is done by xoring undefined extern variables (e.g. \textit{fault\_2}), which we refer to as \textit{fault variables}, to expressions (e.g. to the test condition on line 14).
Eva then treats the fault variable, and thus the entire expression, as potentially having any value.
In the case of nested expressions only the top-level is faulted as this will include the effects of faulting sub-expressions.

\subsection{Dependency Analysis}

Once security properties have been defined and faults simulated we build a dependency graph for the program in order to find the instructions impacting each assertion.
To this end we compute a procedural dependency graph for each function using Frama-C's Pdg plugin \cite{pdgman}.
We then add inter-procedural dependencies by connecting the input and output nodes of these graphs to the corresponding nodes in calls to their respective functions obtained using Frama-C's From plugin.
As Pdg and From are based on data-flow analysis from Eva, the previously simulated faults are taken into account.

%\begin{figure*}[t]
%	\begin{lstlisting}
%extern unsigned int fault_4;
%extern int fault_3;
%extern int fault_2;
%extern size_t fault_1;
%extern unsigned int fault_0;
%
%void print_message(data_t *d_0)
%{
%  /*@ assert rte: mem_access: \valid_read(&d_0->msg_size); */
%  size_t size = (d_0->msg_size & (unsigned int)0xff) ^ fault_4;
%  {
%    size_t i = (unsigned int)0 ^ fault_0;
%    while ((i < size) ^ fault_2) {
%      if ((i > size) ^ fault_3) goto return_label;
%      /*@ assert rte: index_bound: i < 255; */
%      /*@ assert rte: mem_access: \valid_read(&d_0->msg[i]); */
%      printf("%c",(int)d_0->msg[i]); /* printf_va_1 */
%      i = (i + (size_t)1) ^ fault_1;
%    }
%  }
%  printf("\n"); /* printf_va_2 */
%  return_label: return;
%}
%	\end{lstlisting}
%	\caption{Example program with only relevant simulated faults}
%	\label{fig:faults}
%\end{figure*}

All relevant injection points to an assertion are found by exploring the dependency graph starting from its node, selecting those encountered on dependencies.
This way, \textit{fault\_5} on Figure \ref{fig:xor} (line 18) is determined to not have any impact on the assertions on lines 10, 16 and 17 (which were generated using Frama-c's RTE plugin).
We rely on the soundness of Eva's abstract domains computation to ensure that we do not lose dependencies and preserve all attack paths conforming to our fault model.

\begin{figure}[t]
	\centering
	\includegraphics[width=\columnwidth]{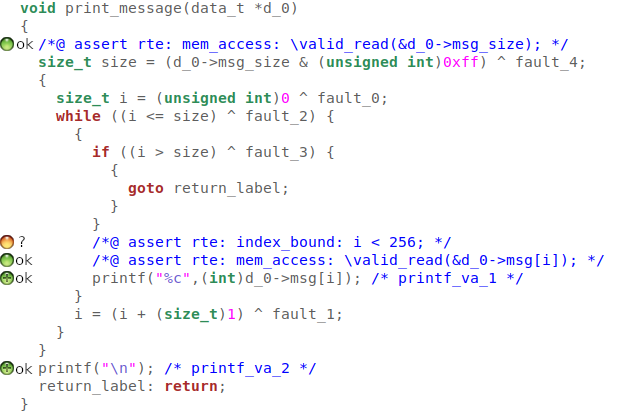}
	\caption{Motivating example with proven and unproven assertions in presence of faults}
	\label{fig:assertex2}
\end{figure}

At this point, we can try to eliminate more injection points by proving that assertions are verified in presence of faults using Eva.
Figure \ref{fig:assertex2} shows which assertions can be proven or not on our motivating example.
Injection points which are only dependencies of proven assertions can be discarded.
Assuming that Eva's proofs are indeed correct, this ensures that no attack path is lost.
Note that Eva assumes that an assertion is true after its annotation (see the second \textit{valid\_read} assertion on Figure \ref{fig:assertex2}).

\subsection{Finding Attack Paths}

Once we have selected injection points, we generate a strategy file for Lazart containing the corresponding fault variables.
In the case of our motivating example we are left with five of them as shown on Figure \ref{fig:strat}.
We also output a source file containing the simulated faults.

\begin{figure}[h!]
	\begin{lstlisting}[numbers=none]
version: 4.0.0
tasks:
  add_trace:
    - __mut__
  rename_bb:
    - __mut__
  countermeasures: []
fault-space:
  functions:
    __all__:
      models:
      - type: data
        vars:
          fault_4: __sym__
          fault_3: __sym__
          fault_2: __sym__
          fault_1: __sym__
          fault_0: __sym__
	\end{lstlisting}
	\caption{Strategy file generated for the previous example program}
	\label{fig:strat}
\end{figure}

We finally run Lazart to find attack paths, targeting the fault variables with the data-load fault model. 
Figure \ref{fig:lazres} shows the results of the analysis, which indicate that only the fault on the masking of the loop bound (line 11 on Figure \ref{fig:xor}) is dangerous in a single fault context.
One way this attack could be fixed would be to check that the index is smaller than 256 since the maximum size of the buffer is fixed.

\begin{figure}[h!]
	\begin{lstlisting}[numbers=none]
Fault Count          0-fault    1-fault
Injection Point
-----------------  ---------  ---------
fault_4                    0          1
	\end{lstlisting}
	\caption{Attacks found by Lazart on our example program}
	\label{fig:lazres}
\end{figure}

While our method mitigates the scalability issues of symbolic execution, it is sometimes not enough as shown in our experimentation.
In such instances results can be obtained by weakening our fault model or eliminating more injection points based on heuristics, which may result in a loss of completeness.
In the case of our motivating example we had to use a fixed value for the size of the message as making it symbolic caused the analysis to be unreasonably slow for such a small program.
In the following section we will discuss a few injection point selection heuristics which can be used to remediate this issue.

\section{Additional Selection Heuristics}

While our method helps to mitigate the scalability issues of symbolic execution, it does not fully eliminate them.
Thus the use of additional selection heuristics is often necessary to further reduce the number of selected injection points and obtain results within a reasonable time frame.

\subsection{Brute-Force Selection}

The brute-force selection approach consists in checking if individual injection points have an impact on the assertions by setting all other fault variables to zero and running Eva, with a timeout mechanism to avoid wasting time when it struggles.
This is effective as only considering one fault improves precision, however this is only valid in a single fault context.
For multi-fault analysis all combinations of injections points would have to be considered, which is not practical as the number of tests that would need to be conducted increases exponentially with the maximum number of faults allowed.

\begin{figure}[t]
    \centering
	\begin{subfigure}{\columnwidth}
		\begin{lstlisting}
unsigned int fault_4 = 0;
int fault_3 = 0;
int fault_2 = 0;
size_t fault_1 = 0;
extern unsigned int fault_0;
		\end{lstlisting}
		\includegraphics[width=\textwidth]{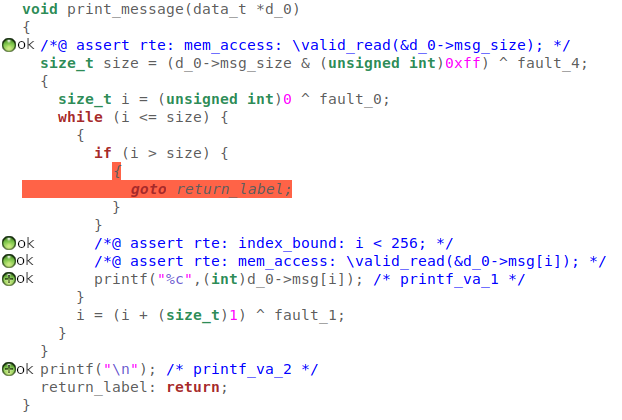}
		\caption{\textit{fault\_0} has no effect}
		\label{fig:f0}
	\end{subfigure}
	\begin{subfigure}{\columnwidth}
		\begin{lstlisting}
extern unsigned int fault_4;
int fault_3 = 0;
int fault_2 = 0;
size_t fault_1 = 0;
unsigned int fault_0 = 0;
		\end{lstlisting}
		\includegraphics[width=\textwidth]{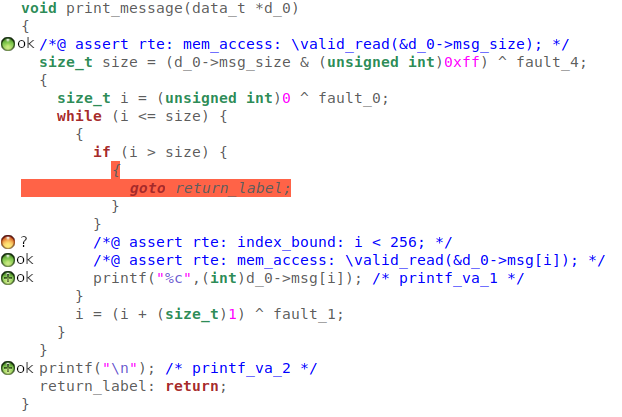}
		\caption{\textit{fault\_4} cannot be proven to have no effect}
		\label{fig:f3}
	\end{subfigure}
	\caption{Example of an injection point with no effect and another which may impact properties}
\end{figure}

On our motivating example, Figure \ref{fig:f0} shows that the \textit{fault\_0} injection point has no effect on the index bound property from line 16 on Figure \ref{fig:xor} (the same is true for \textit{fault\_1}, \textit{fault\_2} and \textit{fault\_3}) while Figure \ref{fig:f3} shows that \textit{fault\_4} may have a negative impact.
This is because single faults targeting the index are caught by countermeasures or restricted by the loop condition while altering the loop bound itself allows for out of bound indexes.
\textit{fault\_4} is therefore the only injection point left to analyze with Lazart and the complexity of the analysis is drastically reduced, allowing us to test with a symbolic message size.
Using the brute-force selection heuristic thus allowed us to ensure completeness in this case.
Note that our plugin removes inactive faults from tests conditions as we noticed they impacted the precision of the analysis.

\subsection{Limiting Fault Injection Point Occurrences}

\begin{figure*}[t]
    \begin{subfigure}{\textwidth}
        \begin{lstlisting}
unsigned int fault_4 = 0;
int fault_3 = 0;
int fault_2 = 0;
size_t fault_1 = 0;
unsigned int fault_0 = 0;

unsigned int fault_0_counter = 0;
unsigned int fault_1_counter = 0;
unsigned int fault_3_counter = 0;
unsigned int fault_2_counter = 0;
unsigned int fault_4_counter = 0;

void print_message(data_t *d_0)
{
  fault_4_counter ++;
  size_t size = (d_0->msg_size & (unsigned int)0xff) ^ fault_4;
  {
    fault_0_counter ++;
    size_t i = (unsigned int)0 ^ fault_0;
    while (1) {
      fault_2_counter ++;
      if (! ((i <= size) ^ fault_2)) 
        break;
      fault_3_counter ++;
      if ((i > size) ^ fault_3) 
        goto return_label;
      printf("%c",(int)d_0->msg[i]);
      fault_1_counter ++;
      i = (i + (size_t)1) ^ fault_1;
    }
  }
  printf("\n");
  return_label: return;
}
        \end{lstlisting}
        \caption{Motivating example with fault occurrence counters}
    \end{subfigure}
    \begin{subfigure}{\textwidth}
        \newcolumntype{Z}{>{\centering}X}
        \centering
        \begin{tabular}{c c c c c c}
            \toprule
            Injection Point & fault\_0 & fault\_1 & fault\_2 & fault\_3 & fault\_4 \\
            Occurrences & 1 & 256 & 257 & 256 & 1 \\
            \botrule
        \end{tabular}
        \caption{Occurrences of each injection point}
    \end{subfigure}
    \caption{Occurrences of injection points in our motivating example}
    \label{fig:fcount}
\end{figure*}

While fault injection points are assimilated to program locations for the purposes of static analysis, during execution these may correspond to multiple instances where faulty behavior may occur.
For example, an single injection point within a loop corresponds to as many potential faults as the maximum number of iterations during execution.

Fault injection points with high occurrence rates are more likely to induce scalability issues as each occurrence generates new paths to explore.
We could thus use an analysis strategy consisting in testing low occurrence injection points first hoping that the analysis terminates in a reasonable amount of time on the first try.
The analysis perimeter can then be expanded by adding higher occurrence injection points until further analysis becomes impractical.

We can obtain occurrence counts during a normal execution by computing the maximum value of counters placed before each injection point using Eva while all faults are inactive.
Figure \ref{fig:fcount} shows our motivating example with injection point occurrence counters and their maximum value.
In this case, the one we showed was dangerous in the previous section, \textit{fault\_4}, only occurs once.
An analysis only over low occurrence injection points would therefore find all single-fault attack paths in this program.

In the case of multi-fault analysis, injection point occurrences should ideally be computed with simulated faults. 
However this yields results too imprecise to be of use in our experience.
The single-fault method can still be used as injection points which are troublesome in single-fault are even more so in multi-fault, however we are missing occurrence data for injection points located in dead code, which may become reachable due to an earlier fault.

We can try to infer "projected" occurrences for these in order to get a selection criteria: when a conditional jump is encountered and one target is dead code, the occurrence count of the jump is attributed to those in the dead code which can be inferred to occur once in an execution of the dead code.  

\begin{figure*}[t]
    \begin{lstlisting}
if(true ^ fault_0){         //n occurrences
    ...
}
else{
    int x = 0 ^ fault_1;    //n projected occurrences
    while(... ^ fault_2){   //no projection
        ...
    }
    x = 1 ^ fault_3;        //n projected occurrences
}
    \end{lstlisting}
    \caption{Example of projected occurrences in dead code}
    \label{fig:proj}
\end{figure*}

Figure \ref{fig:proj} shows a conditional jump with a dead target, occurring n times.
In the dead code \textit{fault\_1} and \textit{fault\_3} have n projected occurrences as they occur once in an isolated execution of that code, while \textit{fault\_2} has no projected occurrences as such execution can loop and visit it multiple times.
We can thus infer that \textit{fault\_1} and \textit{fault\_3} are less likely to cause scalability issues than \textit{fault\_2}.

\subsection{Strategy Shrinking and Growing}

Aside from static analysis, various analysis strategies can be used to detect which injection points cause performance issues during analysis with Lazart.

Strategy shrinking consists in interrupting the analysis after a set amount of time and analyzing KLEE's traces corresponding to unfinished path explorations\footnote{Traces with a .early extension.} in order to find which injection points are triggered the most in them.
These are then assumed to cause performance issues and removed from the strategy file.
This method can be effective if only a few injection points are problematic, however it is difficult to use properly as one need to guess how long the analysis should be running before interrupting. 
It is also not compatible with multi-fault analysis as in that case it is not clear which of the multiple triggered injection points within a trace is causing issues.

Strategy growing is a similar method consisting in adding injection points to the analysis one-by-one, keeping them if the analysis then terminates within the allowed time frame.
This approach is more reliable as previous terminating analyses give a point of reference to set timeouts i.e. an analysis with one more injection point can be expected to terminate within the same time as the previous one plus some margin.
For this reason it is better suited for multi-fault analysis, although the order in which injection points are introduced may impact which ones are kept.
For example, if one injection point is problematic when coupled with n others and is added first, then the n others will be rejected.
However if these n points were added first then only the former may be rejected.
Note that in the case of single fault analysis the injection points can be tested individually to save time.

\subsection{Conclusion}

\begin{table}
    \caption{Characteristics of the Selection Heuristics}
    \label{tab:heur}
    \begin{tabularx}{\columnwidth}{c c c}
        \toprule
        Heuristic & Fault Context & Completeness \\
        \midrule
        Brute-force Selection & single & \ding{51} \\
        Occurrence Limit & single & - \\
        Strategy Shrinking & single & - \\
        Strategy Growing & multi & - \\
        \botrule
    \end{tabularx}
\end{table}

To our knowledge, there is no miracle solution to the scalability issues of symbolic execution, especially when treating it as a black box.
As heuristics are inherently unreliable, maintaining a diverse toolbox of them is key to be efficient.
Table \ref{tab:heur} shows if those we discussed are valid in single- or multi-fault contexts and whether they incur a loss of completeness.
In general the static analysis based ones are faster and should therefore be tried first before additionally resorting to strategy shrinking or growing.

In the next section we will validate our method by testing our implementation in some realistic use-cases.
In particular, we will verify that we do not lose attack paths compared to selecting all injection points in the target program while exploring less executions paths and observing better performance as a result.

\section{Experiments}

\label{sec:res}

We tested our method by analyzing a commonly considered example from the FISSC fault injection test suite \cite{fissc} as well as several real world programs.
In this section we first showcase our method on the verifyPIN example from FISSC in order to illustrate its benefits.
We then present our analysis of the password verification part of the \textit{sudo} unix command from Linux-PAM \cite{sudo}.
Next we analyze the iso7816 library \cite{code_iso} from the ANSSI's WooKey project \cite{wookey} and show that static analysis alone can be very precise in some instances.
Finally, we discuss how our method helped to discover single-fault attack paths bypassing countermeasures in WooKey's bootloader \cite{code_boot}, propose some fixes and test their effectiveness in single- and double-fault contexts.

We used the injection point selection heuristics discussed in the previous section when necessary.
We preferred strategy growing over shrinking, despite it being slower in some cases, as it is less reliant on guesswork and is thus more appropriate in scenarios where evaluators have limited knowledge of their target.

The main criteria we use to evaluate our method are the number of injection points left to analyze with Lazart and the time required to obtain results.
Symbolic execution metrics such as the number of explored paths are also considered.

\subsection{VerifyPIN}

FISSC \cite{fissc} is a collection of programs designed to test fault analysis techniques.
It is mainly composed of variants of verifyPIN, a small PIN code verification program, with different countermeasures applied to it. 
In a single fault context, we analyzed the basic variant without countermeasure and one with Lalande's countermeasure \cite{lalande}, which consists in propagating and checking an instruction counter to reliably detect skips of at least two instructions.
While this countermeasure is not useful against our fault model, which does not allow for violations of the control-flow graph of the program, it adds a lot of complexity to the analysis.

\begin{table}[t]
    \begin{minipage}{\columnwidth}
        \renewcommand*{\thefootnote}{\alph{footnote}}
        \newcolumntype{Z}{>{\centering}X}
        \caption{Results of the analysis of verifyPIN (single faults)}
        \label{tab:vpin}
        \begin{subtable}{\columnwidth}
            \caption{without countermeasures}
            \begin{tabularx}{\columnwidth}{Z Z c c Z c}
                \toprule
                \multicolumn{2}{c}{Static Analysis} & \multirow{2}{*}{IP} & \multicolumn{3}{c}{Lazart Analysis} \\
                Deps & Time & & AP & EP & Time \\
                \midrule
                - & - & 20 & 5 & 65 & 14s \\
                \ding{51} & 1s & 10 & 5 & 52 & 11s \\
                \botrule
            \end{tabularx}
        \end{subtable}
        \begin{subtable}{\columnwidth}
            \caption{with Lalande's countermeasure}
            \begin{tabularx}{\columnwidth}{Z Z c c Z c}
                \toprule
                \multicolumn{2}{c}{Static Analysis} & \multirow{2}{*}{IP} & \multicolumn{3}{c}{Lazart Analysis} \\
                Deps & Time & & AP & EP & Time \\
                \midrule
                - & - & 142 & 6 & 3 253 & 1h 43min \\
                \ding{51} & 1s & 15 & 6 & 1 051 & 30min \\
                \botrule
            \end{tabularx}
        \end{subtable}
        \footnotetext{Deps: Dependencies}
        \footnotetext{IP: Injection Points}
        \footnotetext{AP: Attack Paths}
        \footnotetext{EP: Explored Paths}
    \end{minipage}
\end{table}

Table \ref{tab:vpin} shows the results of our analysis.
As expected, some attacks paths were found where the attacker was able to authenticate with an incorrect PIN code.

As shown in the first part, our method yielded a small improvement in terms of runtime and explored paths during symbolic execution in the case of the basic version of verifyPIN.
The most significant difference lies with the number of injection points considered, which is halved.
This is helpful to evaluators trying to interpret the results of the analysis as only relevant parts of the target have injection points as opposed to everything indiscriminately.

The second part shows that the dependency analysis allowed to eliminate most injection points introduced by Lalande's countermeasure, resulting in the number of injection points to consider being reduced by around $90\%$.
This then induces a greater than $66\%$ reduction in the number of explored paths and the duration of analysis with Lazart, for a negligible cost.

Finally, we find the same number of attack paths in both instances between the two approaches.
Our method thus did not induce a loss of attack paths relatively to our fault model in this case.

\subsection{Sudo}

\begin{table*}[t]
    \begin{minipage}{\textwidth}
        \renewcommand*{\thefootnote}{\alph{footnote}}
        \newcolumntype{Z}{>{\centering}X}
        \caption{Results of the analysis of sudo (single faults)}
        \label{tab:sudo}
        \begin{subtable}{\textwidth}
            \caption{without limits on injection point occurrences}
            \begin{tabularx}{\textwidth}{c c Z Z Z Z c Z c}
                \toprule
                \multirow{2}{*}{Deps} & \multirow{2}{*}{BF} & \multicolumn{2}{c}{Static Analysis} & \multicolumn{2}{c}{Strategy Growing} & \multicolumn{3}{c}{Final Lazart Analysis} \\
                & & IP & Time & IP & Time & AP & EP & Time \\
                \midrule
                - & - & - & - & 919 $\rightarrow$ 913 & 23min & 17 & 737 & 11s \\
                \ding{51} & - & 104 & 1s & 104 $\rightarrow$ 102 & 3min & 17 & 670 & 11s \\
                \botrule
            \end{tabularx}
        \end{subtable}
        \begin{subtable}{\textwidth}
            \caption{with limits on injection point occurrences}
            \begin{tabularx}{\textwidth}{c c Z Z Z Z c Z c}
                \toprule
                \multirow{2}{*}{Deps} & \multirow{2}{*}{BF} & \multicolumn{2}{c}{Static Analysis} & \multicolumn{2}{c}{Occurrences} & \multicolumn{3}{c}{Final Lazart Analysis} \\
                & & IP & Time & IP & Limit & AP & EP & Time \\
                \midrule
                \ding{51} & - & 104 & 1s & 104 $\rightarrow$ 40 & 1 & 10 & 602 & 10s \\
                %\ding{51} & - & 104 & 1s & 104 $\rightarrow$ 47 & 5 & - & - & - \\
            \end{tabularx}
        \end{subtable}
        \footnotetext{Deps: Dependencies}
        \footnotetext{BF: Brute Force (selection heuristic)}
        \footnotetext{IP: Injection Points}
        \footnotetext{AP: Attack Paths}
        \footnotetext{EP: Explored Paths}
    \end{minipage}
\end{table*}

We analyzed the \textit{pam\_sm\_authenticate} function from Linux-PAM, which implements password verification in \textit{sudo}, looking for paths violating the property that one cannot authenticate with a wrong password.
Our goal was to show the performance benefits of our method on a fairly large program (3.6k lines of analyzed code) rather than finding attacks, which was expected given the lack of countermeasures against fault injection.
In this case we had to use strategy growing in order to remove a few injection points causing path explosion issues.

Table \ref{tab:sudo} shows that the dependency analysis allowed to reduce the number of injection points to consider by almost 90\%, which then directly impacts the duration of the strategy growing step as each one is tested separately.
Our approach is thus very beneficial to use this heuristic effectively as it allowed to reduce the analysis time by roughly 85\% in this case, while no attack paths were lost in the end compared to not using it.

We also tested injection points based on their occurrences as shown in the second part of Table \ref{tab:sudo}.
We found that limiting occurrences to one yields results fast, although not all attack paths are discovered.
However increasing the limit results in problematic injection points being added and thus strategy growing would have to be employed to test the remaining ones.

\subsection{Wookey}

Both of our last two targets are components of the ANSSI's WooKey project \cite{wookey}, a secure encrypted USB storage device requiring user authentication in order to access its content.
After attacks were found on it by ITSEFS \cite{inter_cesti}, WooKey was hardened with countermeasures, some against fault injection.
However the effectiveness of these had yet to be tested in the project's current version (0.9).
We thus chose to analyze WooKey's iso7816 \cite{code_iso} and bootloader \cite{code_boot} (2.5k and 3.2k lines of code respectively) in order to check that the attacks had indeed been fixed.
Note that while Lazart has been used during the evaluation of WooKey's Bootloader \cite{inter_cesti}, only the test inversion fault model was considered and the analysis perimeter was set manually. 
In contrast, we attempted to automate the discovery of complex fault attack paths with Lazart using our method.

\subsubsection{Iso7816}

The attack on WooKey's iso7816 library, which implements communication with a security token (a smartcard) containing cryptographic secrets, consisted in using a single fault to modify a loop bound in order to cause repeated buffer overflows similarly to our motivating example.
We thus chose to use memory integrity properties generated by Frama-C's RTE plugin as the starting point of our analysis.
We also chose to disable the test inversion part of our fault model as such faults usually only lead to off-by-one overflows.

\begin{table*}[t]
    \begin{minipage}{\textwidth}
        \renewcommand*{\thefootnote}{\alph{footnote}}
        \newcolumntype{Z}{>{\centering}X}
        \caption{Results of the analysis of WooKey's iso7816 (vulnerable version, single data faults)}
        \label{tab:iso}
        \begin{tabularx}{\textwidth}{Z c c c c c c c c c c}
            \toprule
            \multirow{2}{*}{Deps} & \multirow{2}{*}{BF} & \multicolumn{2}{c}{Assertions} & \multicolumn{2}{c}{Static Analysis} & \multicolumn{2}{c}{Strategy Growing} & \multicolumn{3}{c}{Final Lazart Analysis} \\
            & & Total & Unproven & IP & Time & IP & Time & AP & EP & Time \\
            \midrule
            - & - & - & - & - & - & 660 $\rightarrow$ 655 & 16min & 1 & 192 & 19s \\
            \ding{51} & - & 384 & 3 & 153 & 3s & 153 $\rightarrow$ 151 & 6min & 1 & 170 & 12s \\
            \ding{51} & \ding{51} & 384 & 3 & 1 & 55s & - & - & 1 & 45 & 1s \\
        \end{tabularx}
        \footnotetext{Deps: Dependencies}
        \footnotetext{BF: Brute Force (selection heuristic)}
        \footnotetext{IP: Injection Points}
        \footnotetext{AP: Attack Paths}
        \footnotetext{EP: Explored Paths}
    \end{minipage}
\end{table*}

As we found no attack path with a single fault on the current version of iso7816, we reverted the originally vulnerable part of the code, in the \textit{SC\_get\_ATR} function, back to its previous state.
This allowed us to find the original attack as shown on Table \ref{tab:iso}.
While the dependency analysis alone allowed to reduce the number of injection points considered by over 75\%, the brute force selection heuristic managed to single out the injection point responsible for the violation of the three remaining properties and the attack as the relative simplicity of the properties in relation to the code allowed Eva to be very precise.
This is illustrated by the fact that over 99\% of the assertion could be eliminated.

For this example we again used the strategy growing method in order to obtain results within a controlled time frame.
Testing injection points based on occurrences did not help in this case as all of them were very similar.
The resulting analysis times showcase the importance of reducing the number of injection points to consider as much as possible, as each one had to be individually checked.

\subsubsection{Bootloader}

The attack that was found originally on WooKey's bootloader used a single fault to cause an outdated version of the firmware to be booted.
As WooKey uses a dual-bank system allowing to store two firmwares (flip and flop) so that the older one can be overwritten while the other one continues to operate during updates, this attack was due to the firmware selection logic being unprotected against fault injection.

\begin{table*}[t]
    \begin{minipage}{\textwidth}
        \renewcommand*{\thefootnote}{\alph{footnote}}
        \newcolumntype{Z}{>{\centering}X}
        \caption{Results of the analysis of WooKey's bootloader (single faults)}
        \label{tab:boot}
        \begin{subtable}{\textwidth}
            \caption{without limits on injection point occurrences}
            \begin{tabularx}{\textwidth}{c c Z Z Z Z c Z c}
                \toprule
                \multirow{2}{*}{Deps} & \multirow{2}{*}{BF} & \multicolumn{2}{c}{Static Analysis} & \multicolumn{2}{c}{Strategy Growing} & \multicolumn{3}{c}{Final Lazart Analysis} \\
                & & IP & Time & IP & Time & AP & EP & Time \\
                \midrule
                - & - & - & - & 485 $\rightarrow$ 474 & 19min & 12 & 40 246 & 53min \\
                \ding{51} & - & 226 & 2s & 226 $\rightarrow$ 215 & 15min & 12 & 38 526 & 45min \\
                \ding{51} & \ding{51} & 45 & 3min & 45 $\rightarrow$ 41 & 5min & 12 & 13 592 & 6min \\
                \botrule
            \end{tabularx}
        \end{subtable}
        \begin{subtable}{\textwidth}
            \caption{with limits on injection point occurrences}
            \begin{tabularx}{\textwidth}{c c Z Z Z Z c Z c}
                \toprule
                \multirow{2}{*}{Deps} & \multirow{2}{*}{BF} & \multicolumn{2}{c}{Static Analysis} & \multicolumn{2}{c}{Occurrences} & \multicolumn{3}{c}{Final Lazart Analysis} \\
                & & IP & Time & IP & Limit & AP & EP & Time \\
                \midrule
                \ding{51} & - & 226 & 2s & 226 $\rightarrow$ 119 & 1 & 12 & 3 005 & 19s \\
                %\ding{51} & - & 226 & 2s & 226 $\rightarrow$ 121 & 5 & 12 & 3 226 & 22s \\
                \ding{51} & - & 226 & 2s & 226 $\rightarrow$ 129 & 10 & 12 & 4 518 & 36s \\
                %\ding{51} & - & 226 & 2s & 226 $\rightarrow$ 158 & 15 & 12 & 18 873 & 8min \\
                %\ding{51} & - & 226 & 2s & 226 $\rightarrow$ 159 & 30 & 12 & 19 189 & 8min \\
                \ding{51} & - & 226 & 2s & 226 $\rightarrow$ 161 & 50 & 12 & 23 664 & 17min \\
                %\ding{51} & - & 226 & 2s & 226 $\rightarrow$ 175 & 100 & - & - & - \\
                \midrule
                \ding{51} & \ding{51} & 45 & 3min & 45 $\rightarrow$ 24 & 1 & 12 & 717 & 4s \\
                %\ding{51} & \ding{51} & 45 & 3min & 45 $\rightarrow$ 24 & 5 & 12 & 717 & 4s \\
                \ding{51} & \ding{51} & 45 & 3min & 45 $\rightarrow$ 25 & 10 & 12 & 1 111 & 8s \\
                %\ding{51} & \ding{51} & 45 & 3min & 45 $\rightarrow$ 38 & 15 & 12 & 9 345 & 2.5min \\
                %\ding{51} & \ding{51} & 45 & 3min & 45 $\rightarrow$ 38 & 30 & 12 & 9 345 & 2.5min \\
                \ding{51} & \ding{51} & 45 & 3min & 45 $\rightarrow$ 38 & 50 & 12 & 9 345 & 2.5min \\
                \ding{51} & \ding{51} & 45 & 3min & 45 $\rightarrow$ 42 & 100 & 12 & 20 863 & 11min \\
            \end{tabularx}
        \end{subtable}
        \footnotetext{Deps: Dependencies}
        \footnotetext{BF: Brute Force (selection heuristic)}
        \footnotetext{IP: Injection Points}
        \footnotetext{AP: Attack Paths}
        \footnotetext{EP: Explored Paths}
    \end{minipage}
\end{table*}

Despite countermeasures being added consisting in doubling tests in critical functions, our analysis\footnote{The analyzed code is available as part of FISSC \cite{src}.} shows that attack paths still exist as presented on Table \ref{tab:boot}.

\paragraph{Performance Discussion}

As we were only interested in the ``no firmware rollback" property\footnote{``The most recent firmware is booted or an error / security breach is detected."} there were no discharged assertions during this analysis.
However the dependency analysis and especially the brute force selection heuristic were able to eliminate many injection points, resulting in the total duration of the full analysis being reduced by as much as $80\%$ as shown in the first part of Table \ref{tab:boot}.

Interestingly, all of the dangerous injection points found only occur once.
This is shown in the second part of Table \ref{tab:boot} where all $12$ attack paths are found regardless of the limit put on injection point occurrences.
This means that all attack paths could be found in as low as $21$ seconds\footnote{With no brute-force and a occurrence limit of one.}, although there would be no indicator that there are no more at that point.

A few injection points were eliminated in every analysis, all of which are located within a countermeasure.
We can thus assume that they have no impact alone.

\paragraph{Attack Paths Discussion}

Our results allowed us to identify two attacks which look feasible in practice, both regrouping multiple possible paths, for a total of six attack paths among the twelve found.
Other paths were dismissed due to being overly unrealistic, e.g. requiring to directly alter the boot function pointer, or serving no purpose to the attacker, e.g. allowing to switch to a special mode which can be trivially engaged by pressing on a button on the device.

The first attack consists in exploiting logic in the \textit{loader\_exec\_req\_selectbank} function, which decides which firmware to boot.
Figure \ref{fig:selectbank} shows a simplified version of this function.
Inverting the test on line 3 when both flip and flop are bootable results in execution carrying on to the test on line 7 which only checks if the latter can be booted, assuming that one firmware at least cannot.
This leads to flop being selected regardless of its version.
To fix this attack, we propose to also check that flip is not bootable in that test as well as that flop is not bootable in the next one, corresponding to the highlighted text on Figure \ref{fig:selectbank}.

%\begin{figure*}[t]
%	\begin{lstlisting}
%static loader_request_t loader_exec_req_selectbank(loader_state_t nextstate){
%    //...
%    if ((flip_shared_vars.fw.bootable == FW_BOOTABLE && flop_shared_vars.fw.bootable == FW_BOOTABLE) &&
%        !(flip_shared_vars.fw.bootable != FW_BOOTABLE || flop_shared_vars.fw.bootable != FW_BOOTABLE)){
%        //...
%    }
%    if (flop_shared_vars.fw.bootable == FW_BOOTABLE){
%        if(!(flop_shared_vars.fw.bootable == FW_BOOTABLE))
%            goto err;
%        ctx.boot_flop = sectrue;
%        //...
%    }
%    if (flip_shared_vars.fw.bootable == FW_BOOTABLE){
%        ctx.boot_flip = sectrue;
%        //...
%    }
%    //...
%}
%	\end{lstlisting}
%	\caption{Excerpt from the loader\_exec\_req\_selectbank function in WooKey's bootloader \cite{code_boot}}
%	\label{fig:selectbank}
%\end{figure*}

\begin{figure*}[t]
	\begin{lstlisting}
static loader_request_t loader_exec_req_selectbank(loader_state_t nextstate){
    //...
    if ((flip_shared_vars.fw.bootable == FW_BOOTABLE && flop_shared_vars.fw.bootable == FW_BOOTABLE) &&
        !(flip_shared_vars.fw.bootable != FW_BOOTABLE || flop_shared_vars.fw.bootable != FW_BOOTABLE)){
        //...
    }
    if (flop_shared_vars.fw.bootable == FW_BOOTABLE 
        è&&èà flip_shared_vars.fw.bootable != FW_BOOTABLEà) {
        if(!(flop_shared_vars.fw.bootable == FW_BOOTABLE 
            è&&èà flip_shared_vars.fw.bootable != FW_BOOTABLEà))
            goto err;
        ctx.boot_flop = sectrue;
        //...
    }
    if (flip_shared_vars.fw.bootable == FW_BOOTABLE 
        è&&èà flop_shared_vars.fw.bootable != FW_BOOTABLEà){
        ctx.boot_flip = sectrue;
        //...
    }
    //...
}
	\end{lstlisting}
	\caption{loader\_exec\_req\_selectbank function in WooKey's bootloader \cite{code_boot} (with \textit{\textbf{\color{blue}fixes}})}
	\label{fig:selectbank}
\end{figure*}

The second attack takes advantage of a lack of countermeasures in the \textit{loader\_exec\_req\_flashlock} function, which computes the pointer to the boot function of the chosen firmware.
Figure \ref{fig:flashlock} shows a simplified version of this function.
A fault can be used to invert the test on line 4 and boot flip instead of flop.
Simply doubling this test as shown with the highlighted text should be enough to solve this issue in a single fault context.

%\begin{figure*}[t]
%	\begin{lstlisting}
%static loader_request_t loader_exec_req_flashlock(loader_state_t nextstate){
%    //...
%    else if (ctx.dfu_mode == secfalse) {
%        if (ctx.boot_flip == sectrue) {
%            //...
%            ctx.next_stage = (app_entry_t)(FW1_START);
%        }
%    //...
%    }
%}
%	\end{lstlisting}
%	\caption{Excerpt from the loader\_exec\_req\_flashlock function in WooKey's bootloader \cite{code_boot}}
%	\label{fig:flashlock}
%\end{figure*}

\begin{figure*}[t]
	\begin{lstlisting}
static loader_request_t loader_exec_req_flashlock(loader_state_t nextstate){
    //...
    else if (ctx.dfu_mode == secfalse) {
        if (ctx.boot_flip == sectrue) {
            àif (ctx.boot_flip != sectrue)à
                àgoto err;à
            //...
            ctx.next_stage = (app_entry_t)(FW1_START);
        }
    //...
    }
}
	\end{lstlisting}
	\caption{loader\_exec\_req\_flashlock function in WooKey's bootloader \cite{code_boot} (with \textit{\textbf{\color{blue}fixes}})}
	\label{fig:flashlock}
\end{figure*}

\paragraph{Verifying our Fixes}

\begin{table*}[t]
    \begin{minipage}{\textwidth}
        \renewcommand*{\thefootnote}{\alph{footnote}}
        \newcolumntype{Z}{>{\centering}X}
        \caption{Results of the analysis of Wookey's bootloader with fixes}
        \label{tab:bootfix}
        \begin{tabularx}{\textwidth}{c c Z Z c Z Z c c}
            \toprule
            \multirow{2}{*}{Fault Model} & \multirow{2}{*}{Fault Number} & \multicolumn{2}{c}{Static Analysis} & \multicolumn{2}{c}{Strategy Growing} & \multicolumn{3}{c}{Final Lazart Analysis} \\
            & & IP & Time & IP & Time & AP & EP & Time \\
            \midrule
            Full & 1 & 230 & 2s & 230 $\rightarrow$ 219 & 15min & 6 & 38 548 & 46min \\
            Full & 2 & 145\footnotemark[1] & 2s & 145 $\rightarrow$ 142\footnotemark[2] & 30h & 735 & 113 559 & 19min \\
            Test Inversion & 1 & 77 & 2s & 77 $\rightarrow$ 75 & 3.5min & 4 & 11 724 & 4min \\
            Test Inversion & 2 & 77 & 2s & 77 $\rightarrow$ 60 & 1.5h & 274 & 25 517 & 2min \\
            Test Inversion & 2 & 77 & 2s & 77 $\rightarrow$ 59\footnotemark[3] & - & 6 & 24 907 & 2min \\
        \end{tabularx}
        \footnotetext{IP: Injection Points}
        \footnotetext{AP: Attack Paths}
        \footnotetext{EP: Explored Paths}
        \footnotetext[1]{only injection points occurring once with projection}
        \footnotetext[2]{done on a cluster}
        \footnotetext[3]{without \textit{fault\_82}}
    \end{minipage}
\end{table*}

In order to ensure that our recommendations are valid, we applied our patch and tested it both in single and double fault contexts.
As shown on Table \ref{tab:bootfix} our fixes prevent the six previously identified dangerous attack paths in single-fault.
However since all countermeasures on WooKey's bootloader are designed to protect in this context only, allowing a maximum of two faults should result in them being bypassed.
Unfortunately multi-fault analysis is impractical in part due to the strength of our fault model.
To illustrate this, we selected only injection points occurring once with projection and ran a strategy growing analysis on a cluster, which took around 30 hours to complete.
Given that the final analysis is very incomplete, such runtime is unreasonable in most settings.

One way to solve this issue is to restrict our fault model to test inversion faults only.
This drastically reduces the complexity of the analysis while inducing a loss of attack paths, although we can assume that this drawback is limited as all but two of the attack paths found in single-fault correspond to test inversions.
As shown on Table \ref{tab:bootfix} this analysis still yields a large amount of double-fault attack paths while being roughly 20 times faster compared to the cluster analysis.
However most of these paths include a fault on the same injection point, \textit{fault\_82}, used in the four single-fault attack paths, thus they are redundant.
If we remove \textit{fault\_82} we see that only six attack paths remain, which correspond to those we patched in single fault and thus our countermeasures can be bypassed in two faults.

\subsection{Limits}
\label{sec:lim}

Using our method can present some challenges to the user.
In general, expressing properties can be difficult with limited knowledge of the code, as well as determining which ones may be relevant targets for an attacker.
Parameterizing Eva in order to be able to prove these properties is also tricky and increasing precision has a significant impact on the runtime of analyses.
However the dependency analysis works well regardless of precision and is already helpful.

We also designed our method around the fact that our fault model does not allow for control-flow violations, which makes more sense at source level.
Although, our method could be used with other fault models allowing for localized control-flow violations such as chaining \textit{then} and \textit{else} blocks after conditional jumps or skipping function calls.

Finally, multi-fault analysis remains difficult when many injection points cannot be eliminated but can be done with some concessions as we showed with our analysis of WooKey's bootloader.

\subsection{Using our Method to analyze Cryptographic Implementations}

Our method is focused on the analysis of large programs with many complex feasible execution paths rather than cryptographic code.
We already discussed how these two analysis targets fundamentally differ, however we reckon that the idea of applying the main principles of our method in a cryptographic context should be explored.

The main hurdle with analyzing cryptographic code with our method is that the principles of confusion and diffusion in effet in this context tend to render any kind of dependency analysis pointless, as everything should depend on everything else.
As a result the selection of injection points based on dependency becomes equivalent to naive systematic selection.
In practice, we were unable to eliminate any injection points on neither implementations of CRT-RSA nor AES from the FISSC benchmark \cite{fissc}.

Arguably, we find that the idea of limiting the number of injection points through static analysis to improve scalability is not as helpful for analyzing cryptographic implementations.
Indeed such programs tend to be smaller and to adhere to well-known schemes, meaning that it is easier for experts to decide where faults should be injected in this case.
Beating human judgement in this matter would therefore be more difficult with an automated analysis.
Additionally the scalability issues encountered by tools such as Lazart in a cryptographic context are not only related to the number of possible execution paths but also to the complexity of the path constraints.
We also note that Lazart can analyze the CRT-RSA implementation from FISSC without any issues using the state-of-the-art "set data to zero" fault model, including with the Aumuller and Shamir countermeasures \cite{qasa}, thus optimization is not always needed.
It is only when attempting analysis with unconstrained symbolic faults that difficulties arise.

\section{Related Works}

\subsection{Static Analysis}

In Christofi et al. \cite{christofi} the authors attempted to prove the robustness of a CRT-RSA implementation against fault attacks using formal methods.
In particular, they used Frama-C's Eva and WP plugins to prove security properties on a mutated program with simulated faults.
Since their work was focused on their target specifically, they did not tackle issues that would arise when generalizing their method, namely scalability when considering realistically sized programs.
Additionally, as they were only interested in formal proofs, their approach lacks the versatility required to be usable in other contexts, such as to aid with attack path detection via symbolic execution.

Our approach is similar to that used in the SANTE plugin for Frama-C \cite{sante}, which uses static analysis to generate tests with alarms in order to detect runtime errors, but in the context of fault injection.
SANTE is not a dedicated fault analysis tool and uses regular slicing in order to reduce the size of the tests, which makes it unfit for that particular purpose for the reasons we discussed in Section \ref{sec:sli}.
Fault analysis may also require the generation of impractically large amounts of tests depending on the chosen fault models, which is not an issue when using symbolic execution.

Despite the existence of many fault analysis tools, including a few using symbolic execution, reducing the number of fault injection points to be considered in order to improve their scalability and tackle large targets with many possible execution paths is to our knowledge a novel approach.

\subsection{High-level Fault Analysis Tools}

Larsson et al. \cite{Larsson} first proposed to simulate faults in Java applications by injecting symbolic bits at specified memory addresses.
Their implementation involves manually instrumenting the code to indicate where injection points should be placed.
This method is fundamentally similar to Lazart, with a focus on fault tolerance.

SymPLFIED \cite{patta} uses a single symbolic variable to propagate the effects of fault injection using propagation rules and model checking to find attacks. 
Contrary to Lazart, SymPLFIED's approach introduces dangerous paths which are false positives (i.e fault injection that do not produces crashes or violations of security properties).
Furthermore model checking is also sensitive to path space explosion and thus suffers from the same scalability issues as symbolic execution.

ProFIPy \cite{profipy} simulates faults in python programs according to a user-defined fault model by matching code patterns and replacing them with faulty ones. 
Mutants are thus generated for further testing.
This work also focuses on fault tolerance but our method could still be used with it.

Le et al. \cite{le} use symbolic execution at LLVM IR level to find faulty execution paths in C programs, with faults being simulated by intrucing symbolic bits.
This is the closest available alternative to Lazart, although it is mainly designed to assert fault tolerance.

None of the previously mentionned tools support multi-fault analysis.
Lazart is to our knownledge the only available state-of-the-art option for the purpose of high-level fault analysis of C programs with a focus on fault injection attacks and the ability to find attack paths with multiple simulated faults.
Regardless, we argue that the previously discussed tools are similar enough to be compatible with implementations of our method in their respective languages.

\section{Conclusion}

As the need for security evaluation of not only cryptography but also critical algorithms such as authentication, bootloader and firmware update logic in embedded systems grows, so does the need for tools allowing auditors to verify their intuitions, experiment with various properties of their targets and evaluate countermeasures.
These tools allow to save significant amounts of time when analyzing programs with limited insight on their inner workings.
In this work we showed some solutions allowing to approach large applications, where attack paths tend to be non-trivial and can be obscured by incomplete countermeasures, using automated tools widely considered impractical in this context.

Future works could improve the links between static analysis and dynamic symbolic execution.
Issues with non-terminating symbolic execution and multi-fault analysis should also be addressed further.
Finally, our approach could be applied to other tools such as fault simulators.
It could also be applied at binary level by performing the static analysis part at that level or a hybrid approach could be adopted using code analysis to reduce the complexity of binary analysis.

%\begin{appendices}

%\section{Section title of first appendix}\label{secA1}

%An appendix contains supplementary information that is not an essential part of the text itself but which may be helpful in providing a more comprehensive understanding of the research problem or it is information that is too cumbersome to be included in the body of the paper.

%%=============================================%%
%% For submissions to Nature Portfolio Journals %%
%% please use the heading ``Extended Data''.   %%
%%=============================================%%

%%=============================================================%%
%% Sample for another appendix section			       %%
%%=============================================================%%

%% \section{Example of another appendix section}\label{secA2}%
%% Appendices may be used for helpful, supporting or essential material that would otherwise 
%% clutter, break up or be distracting to the text. Appendices can consist of sections, figures, 
%% tables and equations etc.

%\end{appendices}

%%===========================================================================================%%
%% If you are submitting to one of the Nature Portfolio journals, using the eJP submission   %%
%% system, please include the references within the manuscript file itself. You may do this  %%
%% by copying the reference list from your .bbl file, paste it into the main manuscript .tex %%
%% file, and delete the associated \verb+\bibliography+ commands.                            %%
%%===========================================================================================%%

\bibliographystyle{plain}
\bibliography{fdep2021.bib}% common bib file
%% if required, the content of .bbl file can be included here once bbl is generated
%%\input sn-article.bbl

%% Default %%
%%\input sn-sample-bib.tex%

\end{document}